\documentclass[preprint,12pt,sort&compress]{elsarticle}

\usepackage{paralist}
\usepackage{hyperref}
\hypersetup{
    colorlinks=true,
    linkcolor=blue,
    filecolor=magenta,      
    urlcolor=cyan,
}
\usepackage{amssymb,subcaption,amsmath,float}
\usepackage[linesnumbered,ruled,vlined]{algorithm2e}
\usepackage{xcolor}

\SetCommentSty{mycommfont}
\journal{Computers and Geosciences}
\date{}

\usepackage{subcaption}
\usepackage{listings}
\definecolor{codegreen}{rgb}{0,0.6,0}
\definecolor{codegray}{rgb}{0.5,0.5,0.5}
\definecolor{codepurple}{rgb}{0.58,0,0.82}
\definecolor{backcolour}{rgb}{1,1,1}

\lstdefinestyle{mystyle}
{
  backgroundcolor=\color{backcolour},   commentstyle=\color{codegreen},
  keywordstyle=\color{magenta},
  numberstyle=\tiny\color{codegray},
  stringstyle=\color{codepurple},
  basicstyle=\ttfamily\footnotesize,
  breakatwhitespace=false,         
  breaklines=true,                 
  captionpos=b,                    
  keepspaces=true,                 
  numbers=left,                    
  numbersep=5pt,                  
  showspaces=false,                
  showstringspaces=false,
  showtabs=false,                  
  tabsize=2
}

\begin{document}

\begin{frontmatter}

\title{Bayesian inference and Markov chain Monte Carlo based estimation of a geoscience model parameter}

\author[inst1]{Saumik Dana}
\author[inst2]{Karthik Reddy Lyathakula}

\affiliation[inst1]{organization={University of Southern California},
            city={Los Angeles},
            postcode={90007}, 
            state={CA},
            country={USA}}
            
\affiliation[inst2]{organization={North Carolina State University},
            city={Raleigh},
            postcode={27607}, 
            state={NC},
            country={USA}}

\begin{abstract}
The critical slip distance in rate and state model for fault friction in the study of potential earthquakes can vary wildly from micrometers to few meters depending on the length scale of the critically stressed fault. This makes it incredibly important to construct an inversion framework that provides good estimates of the critical slip distance purely based on the observed acceleration at the seismogram. The framework is based on Bayesian inference and Markov chain Monte Carlo. The synthetic data is generated by adding noise to the acceleration output of spring-slider-damper idealization of the rate and state model as the forward model.
\end{abstract}




\end{frontmatter}

\section{Introduction}
The quantification of fault slip is achieved using the Rate- and State-dependent Friction (RSF) model for friction evolution, which is considered the gold standard for modeling earthquake cycles on faults~\cite{RuiA1983,SchC1989,MarC1998,wei2021synchronization,jia2021effect,valving}. It is given by
\begin{align}
\label{e:ratestate}
\left.\begin{array}{c}
\mu = \mu_0 + A\ln{\left(\frac{V}{V_0}\right)} + B\ln{\left(\frac{V_0\theta}{d_c}\right)}, \\
\frac{d\theta}{dt}= 1-\frac{\theta V}{d_c},
\end{array}\right.
\end{align}
where $V=|d\boldsymbol{d}/dt|$ is the slip rate magnitude, $a=\frac{dV}{dt}$ which we hypothesize is of the same order as recorded by seismograph, $\mu_0$ is the steady-state friction coefficient at the reference slip rate $V_0$, $A$ and $B$ are empirical dimensionless constants, $\theta$ is the macroscopic variable characterizing state of the surface and $d_c$ is a critical slip distance. 
The critical slip distance is the distance over which a fault loses or regains its frictional strength after a perturbation in the loading conditions \cite{palmer1973growth}. In principle, it determines the maximum slip acceleration and radiated energy to such an extent that it influences the magnitude and time scale of the associated stress breakdown process~\cite{scholz2019mechanics}. Regardless of the importance, it is paradoxical that the values of $d_c$ reported in the literature range from a few to tens of microns as determined in typical laboratory experiments~\cite{scholz2019mechanics}, to $0.1-5\,m$ as determined in numerical and seismological estimates based on geophysical observations~\cite{kaneko2017slip}, and further to several meters as determined in high‐velocity laboratory experiments \cite{niemeijer2011frictional}. Moreover, in most numerical simulations of dynamic rupture propagation with prescribed friction laws, $d_c$ is imposed a priori and its value is often assumed to be constant and uniform on the fault plane. Understanding the physics that controls the critical slip distance and explains the gap between observations from experimental and natural faults is thus one of the crucial problems in both the seismology and laboratory communities \cite{ohnaka2003constitutive}. 

With that in mind, we provide a framework in which synthetic earthquake data is used to quantify uncertainty in critical slip distance. 
While the resolution and coupled flow and poromechanics \cite{danaaugmented,danabayesian,danacg,danacmame,danadesign,danaefficient,danafe2,danajcp,danaperformance,danasiam,danasimple,danasystem,danathesis} associated with subsurface activity in the realm of energy technologies and concomitant earthquake quantification is a hot topic, in this work, we focus on the effect of a standard trigonometric perturbation with exponentially decreasing amplitude. In section 2, we explain the spring slider damper idealization to infer the influence of critical slip distance on RSF without recourse to complicated elastodynamic equations. In section 3, we explain the Bayesian inference framework to inversely quantify uncertainty in the estimation of critical slip distance. In section 4, we present the results from our investigation. In section 5, we present conclusions and outlook for future work.
\section{The forward model}
\begin{figure}[htb!]
\centering
\includegraphics[trim={0 0 0 0},clip,scale=0.45]{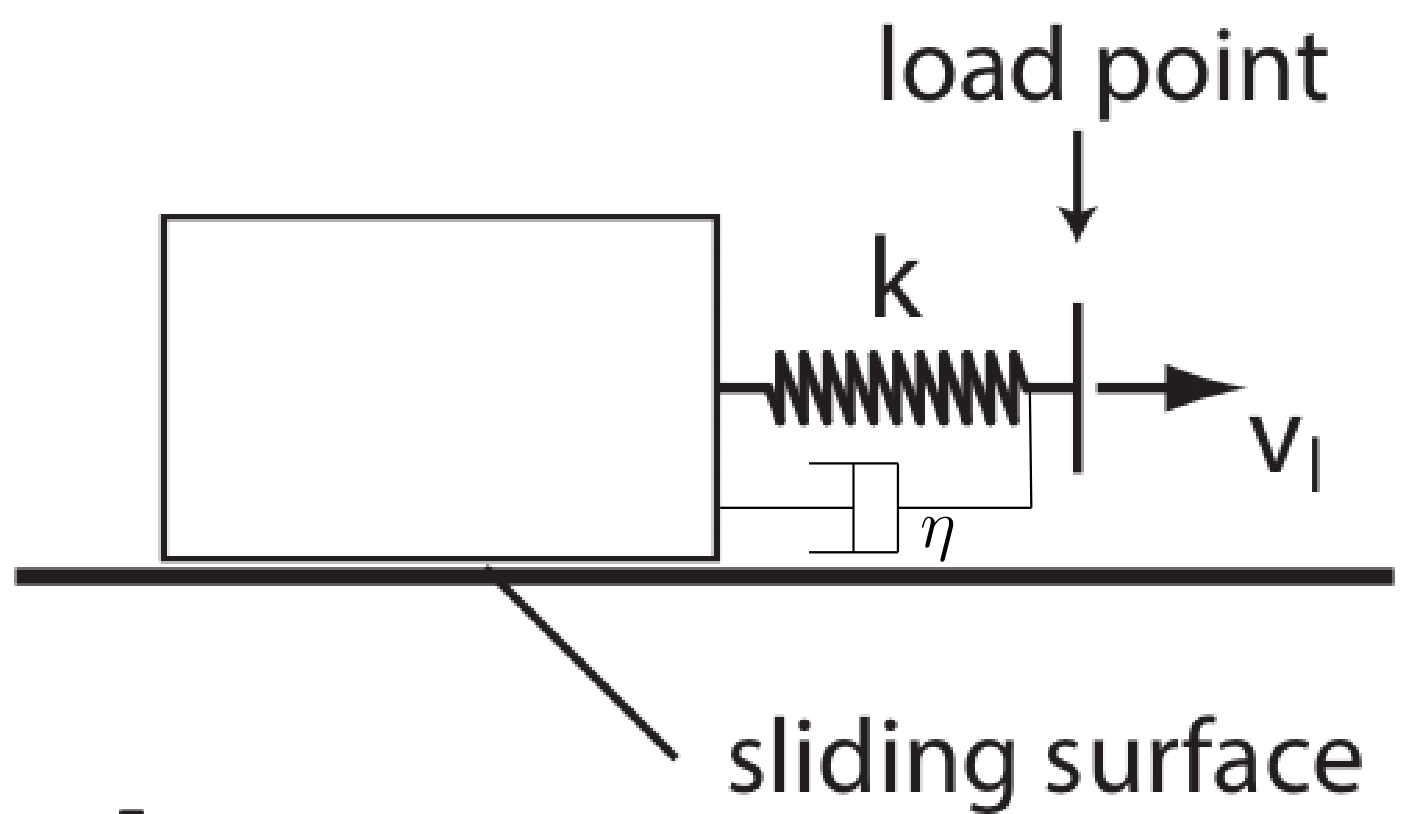}
\caption{Spring Slider Damper Idealization of Fault Behavior}
\label{ss}
\end{figure}
As shown in Fig.~\ref{ss}, we model a fault by a slider spring system \cite{rice1983earthquake,gu1984slip,dieterich1992earthquake}. The slider represents either a fault or a part of the fault that is sliding. The stiffness $k$ represents elastic interactions between the fault patch and the ductile deeper part of the fault, which is assumed to creep at a constant rate. This simple model assumes that slip, stress, and friction law parameters are uniform on the fault patch. The friction coefficient of the block is given by
\begin{equation*}\label{slider_spring1}
\mu= \frac{\tau}{\sigma} = \frac{\tau_l-k\delta-\eta V}{\sigma}
\end{equation*}
where $\sigma$ is the normal stress, $\tau$ the shear stress on the interface, $\tau_l$ is the remotely applied stress acting on the fault in the absence of slip, -$k\delta$ is the decrease in stress due to fault slip \cite{kanamori2004physics} and $\eta$ is the radiation damping coefficient \cite{mcclure2011investigation}. We consider the case of a constant stressing rate $\dot{\tau_l} = kV_l$ where $V_l$ is the load point velocity, which we take to be
\begin{align}
\label{loading}
V_l = V_0(1+\exp{(-t/10)}\,sin(10t))
\end{align}
The stiffness is a function of the fault length $l$ and elastic modulus $E$ as $k\approx \frac{E}{l}$. With $k'=\frac{E}{l\sigma}$, we get
\begin{equation}\label{e:ratestate4}
\dot{\mu} \approx k'(V_l-V)-k''\dot{V}
\end{equation}
where $k''=\frac{\eta}{\sigma}$. Once the phenomenological form of $\dot{\mu}$ is known, we rewrite Eq. \eqref{e:ratestate} as
\begin{equation}\label{e:ratestate1}
\left.\begin{array}{c}
V = V_0\exp\left(\frac{1}{A}\left(\mu - \mu_0 -B\ln{\left(\frac{V_0\theta}{d_c}\right)}\right)\right),\\
\dot{\theta} = 1-\frac{\theta V}{d_c},\\
\dot{V}  = \frac{V}{A}\left(\dot{\mu}-\frac{B}{\theta}\dot{\theta}\right)
\end{array}\right.
\end{equation}
The ballpark values are:
\begin{compactitem}
    \item Elastic modulus $E=5 \times 10^{10}\,Pa$
    \item Critical fault length $l=3\times 10^{-2}\,m$
    \item Normal stress $\sigma=200 \times 10^6 Pa$
    \item Radiation damping coefficient $\eta=20 \times 10^6 Pa/(m/s)$
    \item $A=0.011$
    \item $B=0.014$
    \item $V_0=1 \mu m/s$
    \item $\theta_0=0.6$
    \item $\mu_0=\mu_0=0.6$
\end{compactitem}
from  which the effective stiffness and damping are obtained as
\begin{compactitem}
\item $k'=\frac{E}{l\sigma}=\frac{5 \times 10^{10}}{3\times 10^{-2}\times 2 \times 10^8} [1/m] \approx 10^{-2} [1/\mu m]$
\item $k'' = \frac{\eta}{\sigma} = \frac{2 \times 10^7}{2 \times 10^8} [s/m] = 10^{-7}[s/\mu m]$
\end{compactitem}
\section{Bayesian inversion framework}
\begin{figure}[htb!] 
	\begin{subfigure}{0.5\textwidth}		
		\centering
		\includegraphics[scale=0.45]{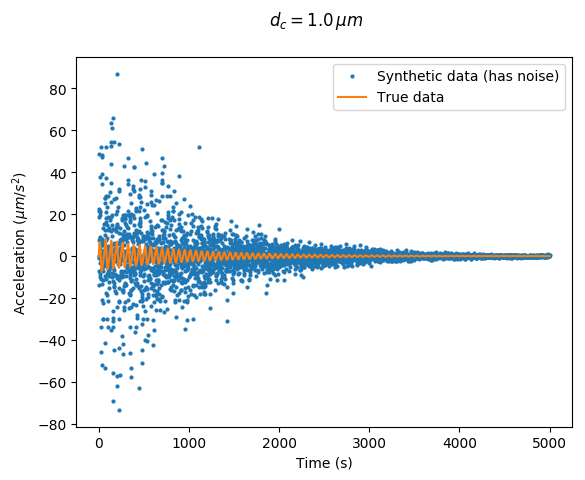}
	\end{subfigure}
	\begin{subfigure}{0.5\textwidth}		
		\centering
		\includegraphics[scale=0.45]{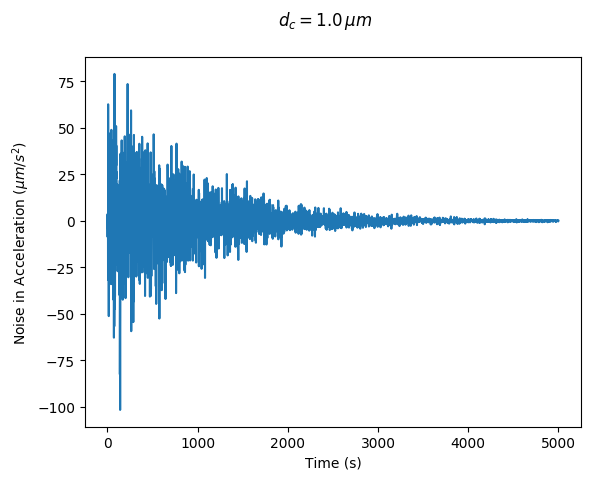}
	\end{subfigure}
	\caption{Noise is modeled as normal distribution}
	\label{response3}
\end{figure}
To define the inverse problem, considered the relationship between acceleration ($a_i(t)$) and the model response by the following statistical model
\begin{equation}\label{e:ratestate_stat}
a(t_i) = f(t_i,\theta,\mu,A,B,d_c)+\epsilon_{i}
\end{equation}
where $\epsilon_{i}$ is the noise. Assuming the $\epsilon_{i} \sim N(0,\sigma^2)$ as unbiased, independent and identical normal distribution with standard deviation $\sigma$ allows us to conveniently generate the synthetic data as shown in Fig.~\ref{response3}. The synthetic data over time $a(t_1),...,a(t_n)$ are the $n$ observations for $a(t_i)$ and $f(A,t_i,\theta,\mu,A,B,d_c)$ is the acceleration response of the model over time obtained using the forward model. The goal of the inverse problem is to determine the model parameter ($d_c$) from the Eq.\eqref{e:ratestate_stat} 
.
Using the Bayes theorem~\cite{smith2013uncertainty}, the distribution for the model parameter is given by the posterior distribution $\pi(d_c|a(t_1),...,a(t_n))$ given by
\begin{equation}\label{e:ratestate_bayes}
\pi(d_c|a(t_1),...,a(t_n))= \frac{\pi(a(t_1),...,a(t_n)|d_c) \pi_0(d_c)}{\int_{d_c} \pi(a(t_1),...,a(t_n)|d_c) \pi_0(d_c) dd_c}
\end{equation}
where $\pi(a(t_1),...,a(t_n)|d_c)$ is the likelihood given by
\begin{align}
\nonumber
\pi(a(t_1),...,a(t_n)|d_c) &=  \prod_{i=1}^{n} \pi(a(t_i)|d_c) \\
\label{e:ratestate_likeli}
&= \prod_{i=1}^{n} \frac{1}{\sigma \sqrt{2\pi}} e^{-\frac{1}{2} \left(\frac{a(t_i)-f(V,t_i,\theta,\mu,A,B,d_c)}{\sigma}\right)^2}
\end{align}
and $\pi_0(d_c)$ is the prior distribution. The denominator is a normalizing integral. The information of the model parameter can be included in the posterior distribution through the prior, $\pi_0(d_c)$. In this study, the prior is assumed to be uniform distribution and the prior is a constant value inside the uniform distribution limits. 
\emph{The quantity that makes evaluation of the posterior difficult is the integral term in the denominator.} Direct evaluation of the integral using quadrature rules is expensive and often requires adaptive methods. Alternatively, sampling methods like Markov chain Monte Carlo (MCMC) methods \cite{shapiro2003monte,hastings1970monte,haario2001adaptive,mueller2010exploring} can be used to evalute the integral. 


The MCMC method generates the parameter samples
from a proposal distribution, and the sample is rejected or accepted by
evaluating the posterior distribution for the corresponding parameter
sample. Adaptive Metropolis MCMC algorithm~\cite{haario2001adaptive} is used as the sampling
method. The MCMC simulation starts from a random value for the model
parameter generated from uniform bounded distribution. The MCMC sampling method uses the obtained predicted acceleration and a new value for the model parameter is generated using the Markov chain. The errors are assumed to be a normal distribution with a standard deviation and at each
iteration of the MCMC sampling method, a new value of standard deviation is generated, associated with the model parameter using inverse-gamma distribution. In the next iteration, the
predictive model is solved again for the new parameter and the obtained
predicted acceleration is used by the MCMC sampling method to generate
the next parameter sample. A total of $n$ MCMC samples are generated from
the MCMC sampling method.

\section{Results}
\begin{figure}[htb!]
\begin{subfigure}{.5\textwidth}
    \centering
    \includegraphics[scale=0.4]{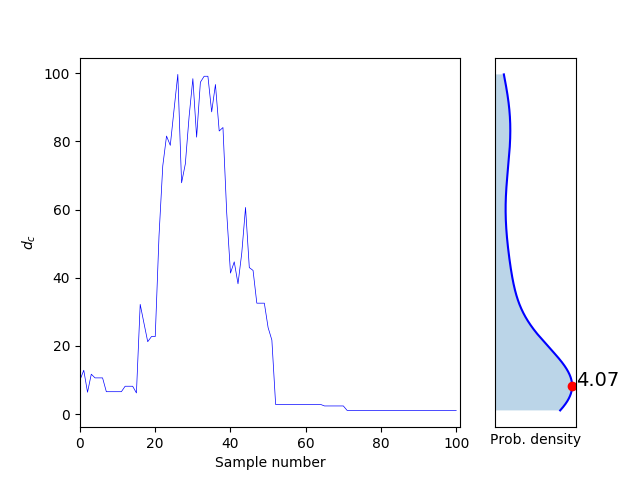}
    \caption{100 samples}
\end{subfigure}
\begin{subfigure}{.5\textwidth}
    \includegraphics[scale=0.4]{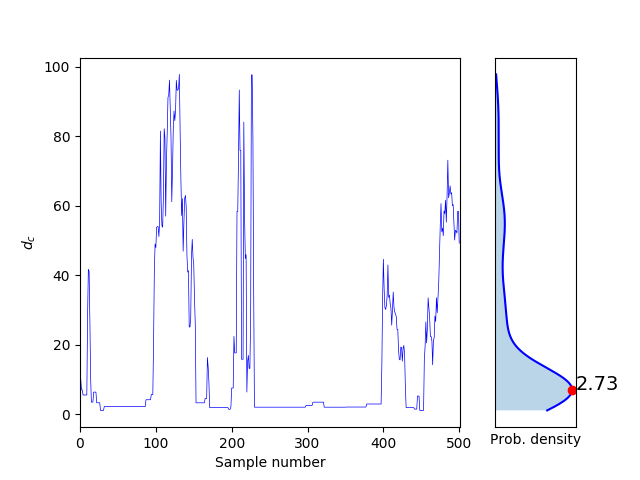}
    \caption{500 samples}
\end{subfigure}
\begin{subfigure}{.5\textwidth}
    \includegraphics[scale=0.4]{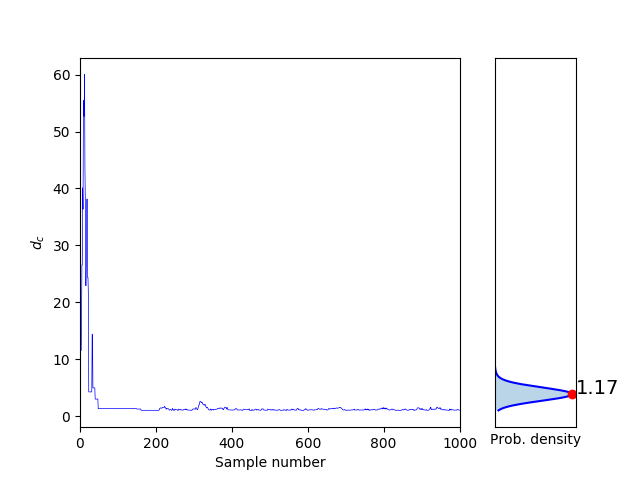}
    \caption{1000 samples}
\end{subfigure}
\begin{subfigure}{.5\textwidth}
    \includegraphics[scale=0.4]{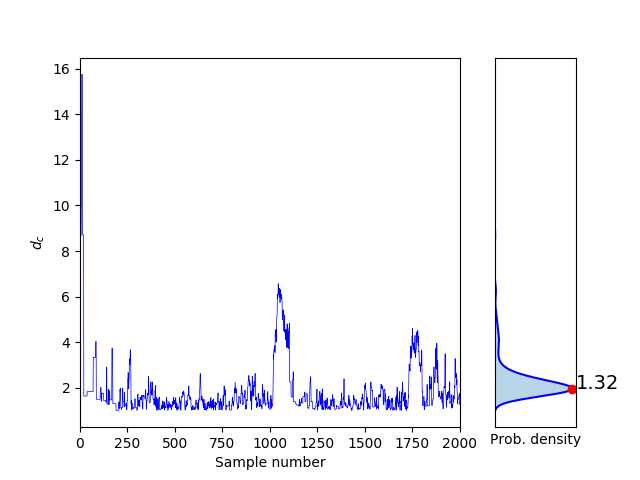}
    \caption{2000 samples}
\end{subfigure}
    \caption{Output of the Bayesian MCMC framework for model parameter for different sample numbers with true value of $d_c=1\,\mu m$ and an initial guess of $10\,\mu m$}
    \label{plot5}
\end{figure}
\begin{figure}[htb!]
\begin{subfigure}{.5\textwidth}
    \centering
    \includegraphics[scale=0.4]{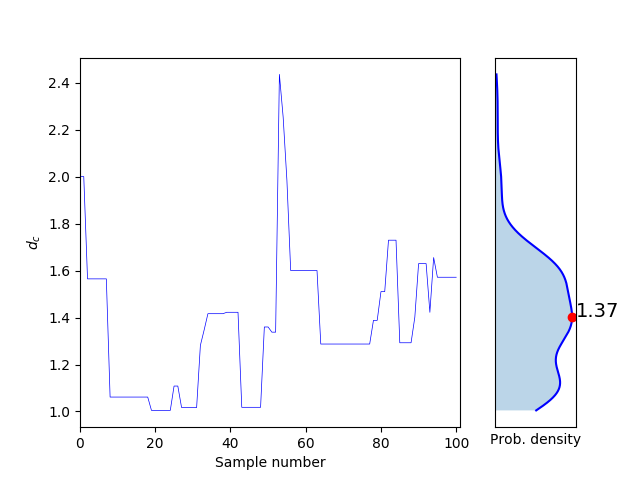}
    \caption{100 samples}
\end{subfigure}
\begin{subfigure}{.5\textwidth}
    \includegraphics[scale=0.4]{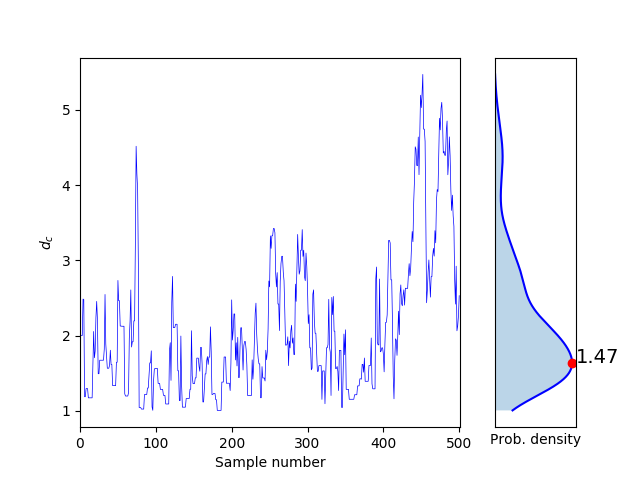}
    \caption{500 samples}
\end{subfigure}
\begin{subfigure}{.5\textwidth}
    \includegraphics[scale=0.4]{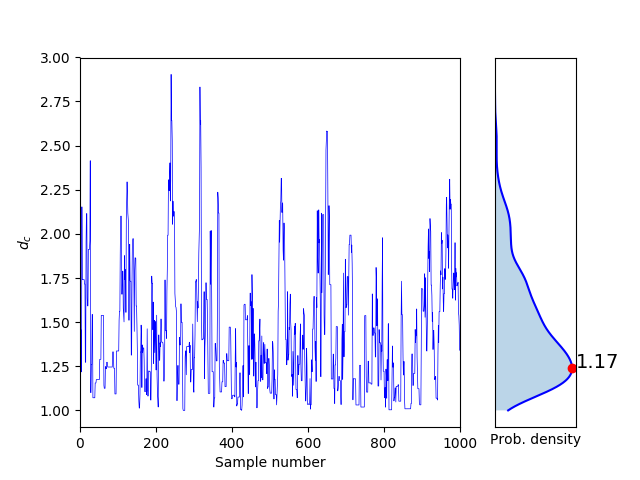}
    \caption{1000 samples}
\end{subfigure}
\begin{subfigure}{.5\textwidth}
    \includegraphics[scale=0.4]{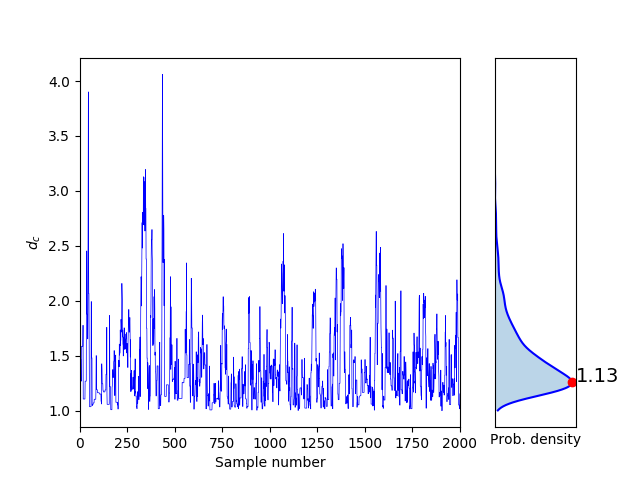}
    \caption{2000 samples}
\end{subfigure}
    \caption{Output of the Bayesian MCMC framework for model parameter for different sample numbers with true value of $d_c=1\,\mu m$ and an initial guess of $2\,\mu m$}
    \label{plot6}
\end{figure}
Figs.~\ref{plot5}-~\ref{plot6} show a gradual improvement in the estimated model parameter as the number of samples is increased. It is evident that the framework has a lot more difficulty converging to the true value if the initial guess is far away from the true value. In reality, it is not really known what the true value is, and arbitrary initial guesses will lend a solution that is way off target regardless of the number of sampling points. We observe that the Bayesian framework settles to a value with maximum probability after a certain number of sampling points regardless of the number of samples following it. In all of the simulations, we observe in all the simulations that the Bayesian framework does an excellent job of dropping to the ballpark of the expected true value within the first few samples. We see that the framework is not heavily reliant on the initial guess for accuracy for the set of simulations we are rendering. 

\section{Conclusions and Outlook}
The rate and state model for fault friction evolution is a piece of the puzzle in which forward simulations are used to arrive at the seismic impact of fault slip on earthquake activity. Typically, the earthquakes are measured with seismograms and geophones on the surface as P-waves and S-waves, and then these readings are used to calibrate the seismic activity for constant monitoring. The acceleration field around the fault slip activity translates to these waves recorded on the surface, and forward simulations with wave propagation bridge that gap. That being said, a seismic recording on the surface cannot easily be backtraced to acceleration around the fault, and finally to the source of the fault slip. That is precisely inverse modeling, and the Bayesian framework coupled with MCMC allows us to put such a framework in place. The accelerations are field quantities, and the inverse estimation of the field around the fault from the time series at the seismogram and/or geophone is not a trivial task. With that in mind, we test the robustness of the Bayesian/MCMC framework to inversely estimate a value instead of a field. The thing is that any inversion framework works on data as the input, and since subsurface data is not available other the sensors at the wells, the forward simulations are used to generate this data with all the computational physics put in place. This generated data is then used to inversely estimate the acceleration field from the data at the surface. Although such forward simulations to generate data are infeasible in real-time scenarios where we need estimates of what is happening in the subsurface from the reading on the surface almost immediately, the framework robustness would eventually lend itself to that scenario. The scenario is that the recording at the seismogram and/or geophone would be fed into the Bayesian/MCMC fraemwork as an input, and the framework would provide an estimate of the acceleration field around the fault as the output. In this work, we work on arriving at estimates of critical slip distance in the rate and state model by running a spring-slider-damper as the forward model instead of using a full-fledged forward simulator. In the future, we will be deploying the coupled flow and geomechanics simulator as the forward model, and gradually arrive at the aforementioned scenario.

\appendix

\bibliographystyle{unsrt}
\bibliography{sample1}

\end{document}